\documentclass[aps,prb,preprint,superscriptaddress,showpacs]{revtex4-1}

\usepackage{graphicx}
\usepackage{amssymb, amsmath}
\usepackage[usenames]{color}
\usepackage{multirow}
\usepackage{url}

\usepackage{subfigure}
\usepackage{epsfig}
\usepackage{dcolumn}
\usepackage{bm}
\usepackage{float}

\begin{document}

\title{Type-II Dirac fermions in the PtSe$_2$ class of transition metal dichalcogenides}
\author{Huaqing Huang}
\affiliation{Department of Physics and State Key Laboratory of Low-Dimensional Quantum Physics, Tsinghua University, Beijing 100084, China}
\affiliation{Collaborative Innovation Center of Quantum Matter, Tsinghua University, Beijing 100084, China}

\author{Shuyun Zhou}
\affiliation{Department of Physics and State Key Laboratory of Low-Dimensional Quantum Physics, Tsinghua University, Beijing 100084, China}
\affiliation{Collaborative Innovation Center of Quantum Matter, Tsinghua University, Beijing 100084, China}

\author{Wenhui Duan\footnote{dwh@phys.tsinghua.edu.cn}}
\affiliation{Department of Physics and State Key Laboratory of Low-Dimensional Quantum Physics, Tsinghua University, Beijing 100084, China}
\affiliation{Collaborative Innovation Center of Quantum Matter, Tsinghua University, Beijing 100084, China}
\affiliation{Institute for Advanced Study, Tsinghua University, Beijing 100084, China}

\date{\today}

\begin{abstract}
Recently, a new ``type-II'' Weyl fermion, which exhibits exotic phenomena such as angle-dependent chiral anomaly, was discovered in a new phase of matter where electron and hole pockets contact at isolated Weyl points. [A. A. Soluyanov, \textit{et al}. Nature \textbf{527}, 495 (2015)]
This raises an interesting question whether its counterpart, i.e., type-II \textit{Dirac} fermion, exists in real materials. Here, we predict the existence of symmetry-protected type-II Dirac fermions in a class of transition metal dichalcogenide materials. Our first-principles calculations on PtSe$_2$ reveal its bulk type-II Dirac fermions which are characterized by strongly tilted Dirac cones, novel surface states, and exotic doping-driven Lifshitz transition. Our results show that the existence of type-II Dirac fermions in PtSe$_2$-type materials is closely related to its structural $P\bar{3}m1$ symmetry, which provides useful guidance for the experimental realization of type-II Dirac fermions and intriguing physical properties distinct from those of the standard Dirac fermions known before.
\end{abstract}


\maketitle

Dirac and Weyl fermions, which are basic concepts in high-energy physics, have attracted much attention in condensed matter physics recently because of the discovery of three-dimensional (3D) Dirac and Weyl semimetals \cite{Na3Bi,Cd3As2,WanXG,Balents,TaAsPRX,TaAsHLin,Jianpeng,CaoWendong}. In Dirac and Weyl semimetals, four-fold and two-fold degenerate point-like linear band crossings, the so-called Dirac and Weyl points, appear in the vicinity of the Fermi level. Hence their low-energy excitations are Dirac and Weyl fermions that obey the Dirac or Weyl equations, and a large variety of novel phenomena such as large linear magnetoresistance and chiral anomaly \cite{ChiralAnomaly,QMR}, have been predicted and observed in Dirac and Weyl semimetals.

Since the restriction of Lorentz invariance is stringent only in high-energy physics whereas not necessary in condensed matter physics, Soluyanov \textit{et al}. \cite{type2Weyl} recently proposed a previously overlooked type of Weyl fermion (type-II) that breaks the Lorentz symmetry in condensed matter systems.
Different from conventional Weyl fermions which have standard Weyl points with point-like Fermi surfaces (which we denote as ``type-I''), the type-II Weyl fermion emerges at the boundary between electron and hole pockets in materials such as WTe$_2$ and MoTe$_2$. Such type-II Weyl fermions have recently been theoretically predicted \cite{type2Weyl,BinghaiMoTe2,*QPIMoTe2,*HsinLinMoWTe2,*ZJWangMoTe2} and experimentally verified by different research groups recently \cite{shuyunMoTe2,*DinghongMoTe2,*ZhouXJMoTe2,*YLChenMoTe2,*huang2016MoTe2,wang2016spectroscopic,*wu2016observation,*bruno2016surface}.
The discovery of type-II Weyl fermions not only is of great importance for basic science, but also brings lots of new physical properties, such as angle-dependent chiral anomaly and topological Lifshitz transitions \cite{type2Weyl,volovik2016lifshitz,YongXu} that are distinct from the type-I Weyl fermions known before. In the course of searching for new fermions, the following questions naturally arose: Is there a Dirac counterpart of the type-II Weyl fermion, namely, a ``type-II Dirac fermion'', in condensed matter systems?

So far, several Dirac semimetal materials, such as Na$_3$Bi \cite{Na3Bi,Na3Bi_exp1,*Na3Bi_exp2,*Na3Bi_exp3} and Cd$_3$As$_2$ \cite{Cd3As2,Cd3As2_exp1,*Cd3As2_exp2,*Cd3As2_exp3}, have been theoretically proposed and experimentally confirmed. However, only conventional type-I Dirac fermions exist in all of these materials. Here, we predict the existence of type-II Dirac fermions in transition metal dichalcogenides PtSe$_2$, PtTe$_2$, PdTe$_2$ and PtBi$_2$ using first-principles calculations and effective Hamiltonian analysis. Taking PtSe$_2$ as a representative, we find that two symmetry-protected type-II Dirac points are located along the $k_z$ axis in the bulk Brillouin zone (BZ). Different from standard type-I Dirac points with point-like Fermi surfaces, the type-II Dirac points in PtSe$_2$ appear at the contact of electron and hole pockets with strongly tilted Dirac cones. Furthermore, novel surface states and exotic doping-driven Lifshitz transition occur in these materials.
We believe that these materials are promising platforms for the experimental realization of the new type-II Dirac fermions and novel physical properties different from those of standard type-I Dirac semimetals.

All first-principles calculations are carried out within the framework of density-functional theory (DFT) using the Perdew-Burke-Ernzerhof-type \cite{PBE} generalized gradient approximation for the exchange-correlation functional, which is implemented in the Vienna \textit{ab initio} simulation package \cite{VASP}. An $8\times8\times8$ grid of \textbf{k} points and a 520 eV plane-wave energy cutoff are adopted for the self-consistent field calculations. Spin-orbit coupling (SOC) is taken into account in our calculations. We construct Wannier functions by projecting the Bloch states from the first-principles calculations of bulk materials onto a set of atomiclike trial orbitals without an iterative maximal-localization procedure \cite{wannier1,*wannier2,wannier90,InterfaceTICI,huanghq_NodalLine}. Based on the Wannier representation, we further calculate the surface spectral function and Fermi surface of the semi-infinite system using a surface Green's function method \cite{lopez,*lopez2,PhysRevB.90.195105,semiDirac}.

\begin{figure}
\includegraphics[width =\columnwidth]{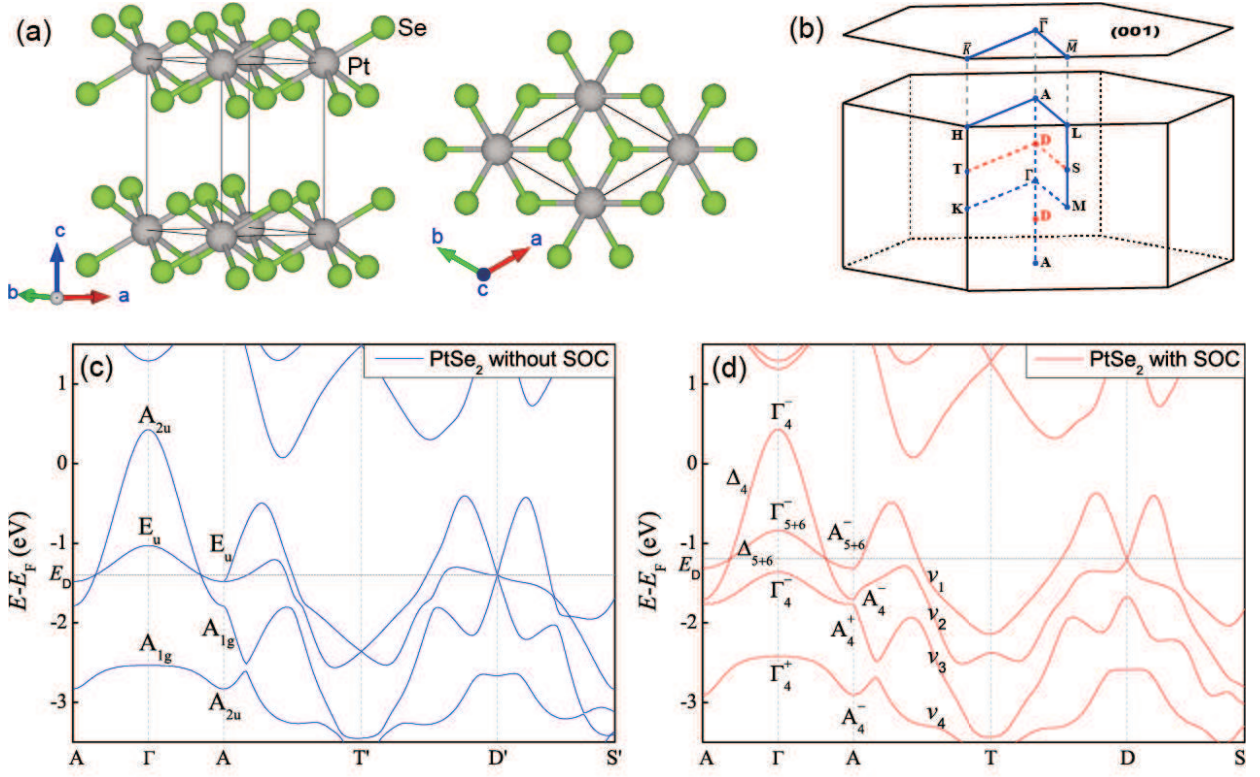}%
\caption{\label{fig1_band} (Color online) Crystal structure and electronic structure of PtSe$_2$. (a) Crystal structure of PtSe$_2$ with $P\bar{3}m1$ symmetry. (b) Brillouin zone (BZ) of bulk and the projected surface BZs of (001) surface. Blue dots indicate the high-symmetry points of the BZ, and red dots highlight the 3D Dirac point position. (c,d) The bulk band structures of PtSe$_2$ with/without SOC. The valence bands are labeled as $v_i (i\!=\!1$-$4)$ in terms of their distance to the Fermi level. $E_D$ is the energy of the Dirac points. The irreducible representations of selected bands along the high symmetric $\mathbf{k}$ path are indicated.}
\end{figure}

As all mentioned materials share similar crystal and electronic structures, we take PtSe$_2$ as a representative material hereafter.
PtSe$_2$ crystalizes in a centrosymmetric trigonal crystal structure with space group $P\bar{3}m1$ (No.~164, $D_{3d}^3$). The calculated lattice constants are $a\!=\!3.784$~\AA~and $c\!=\!5.107$~\AA. The atomic structure of PtSe$_2$ is shown in Fig.~\ref{fig1_band}(a). PtSe$_2$ is a periodic stack of layered basic building block with weak inter-layer interactions. In each layer, Pt atoms are sandwiched by top and bottom Se layers, whereas two Se atoms are related by inversion symmetry. The bulk and surface BZs are shown in Fig.~\ref{fig1_band}(b), where high-symmetry points, lines and Dirac points are also indicated.

We first obtain the band structure of PtSe$_2$ without SOC as shown in Fig.~\ref{fig1_band}(c). There are clear band crossing features near the Fermi level along the $\Gamma$-$A$ line. As the two valence bands belong to one-dimensional (1D) irreducible representation (IR) $A_{2}$ and two-dimensional (2D) $E$, respectively, the two bands can simply cross each other without opening a gap. We also check the band crossing along the in-plane $S$-$D$-$T$ momentum path (see Fig.~\ref{fig1_band}(b)). As shown in Fig.~\ref{fig1_band}(c), the doubly degenerate ${E}$ band split into two bands as the symmetry decreases in this line, hence the band crossing is actually a triply (or sixfold if the spin degree of freedom is considered) degenerate point, which rarely appears in band structures due to the restriction of co-dimension \cite{huanghq_NodalLine} and can be regarded as a new fermion \cite{newfermion,triple1,*triple2}.

Due to the strong SOC of Pt and Se, including SOC in our first-principles calculations leads to a dramatic modification of the electronic structure as shown in Fig.~\ref{fig1_band}(d). In the presence of SOC, originally degenerate $E$ bands along the $\Gamma$-$A$ line split. However, the band crossing still exists with only a small shift of the crossing point. And IRs of the crossing bands become 2D $\Delta_4$, 1D $\Delta_5$ and $\Delta_6$ in the double point group $C_{3v}(3m)$ representation of the $\Gamma$-$A$ line \cite{koster1963properties}. Because the $\Delta_5$ and $\Delta_6$ bands are always degenerate due to the coexistence of time reversal and inversion symmetry, the isolated band crossing is a four-fold degenerate Dirac point. More importantly, the Dirac cone is tilted strongly along $\Gamma$-$A$ but untilted along in-plane lines ($S$-$D$-$T$), which is the characteristic feature of type-II Dirac fermions mentioned above. According to our calculations, a pair of symmetry related Dirac points are located at the $D$ point, $\mathbf{k}^c=(0,0,\pm0.319)$ (in units of $2\pi/c$), on the $k_z$ axis with an energy of $E_D\!=\!-1.193$ eV below the Fermi level. We also calculate the electronic structures of PtSe$_2$ under uniaxial strain along the $c$ axis, which does not violate the trigonal symmetry of the system, and find that it is possible to tune the Dirac point around the Fermi level by combining external strains with other techniques such as chemical doping (see the Supplementary Materials \footnote{\label{fn}See Supplemental Material at http://link.aps.org/supplemental/10.1103/PhysRevB.94.121117 for Type-II Dirac fermions in the PtSe$_2$ class of transition metal dichalcogenides.}).
Remarkably, the band structures of other PtSe$_2$-class materials (i.e., PtTe$_2$, PdTe$_2$ and PtBi$_2$) are very similar. We have calculated the electronic structures and locations of Dirac points for all materials \footnotemark[\value{footnote}]. The precise positions of the Dirac points for these materials are summarized in Table. I.

Interestingly, there is a band inversion in $\Gamma$-$A$ between the third and fourth valence bands (labeled as $v_3$ and $v_4$) with opposite parities, which indicates nontrivial $\mathbb{Z}_2$ topology. We then perform a direct calculation of $\mathbb{Z}_2$ invariant using the Wannier charge center methods \cite{alexey2}, and find $\mathbb{Z}_2\!=\!(1;000)$ for the subspace spanned by valence bands below the $v_3$ band (i.e., the $v_4$  band and bands below it)\cite{pdte2}.

\begin {table}
\caption {The momentum space position (0,0,$\pm k_z^c$) and energy $E_D$ (with respect to the Fermi level) of Dirac points.}
\label{tab:Dirac_position}
\begin{center}
\begin{tabular}{lcc}
  \hline
  \hline
        & $k_z^c$ ($2\pi/c$) & $E_D$ (eV) \\
  \hline
  PtSe$_2$ & 0.319 & -1.193  \\
  PtTe$_2$ & 0.346 & -0.860  \\
  PdTe$_2$ & 0.406 & -0.545  \\
  \multirow{2}{*}{PtBi$_2$\footnote{There are two closely located Dirac points in PtBi$_2$: One is of type-I, and the other is of type-II.}}
 & 0.163 & 2.539 \\
                              & 0.216 & 2.429  \\
  \hline
  \hline
\end{tabular}
\end{center}
\end{table}

\begin{figure}
\includegraphics[width = \columnwidth]{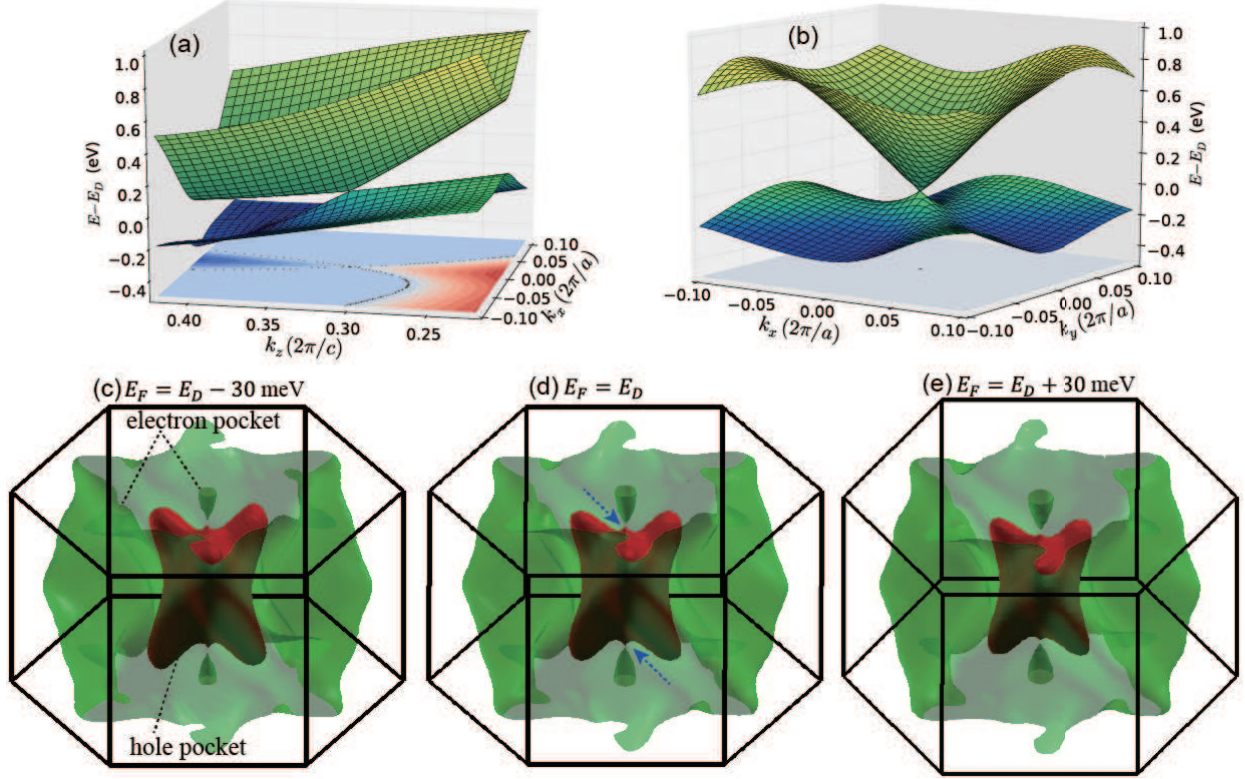}%
\caption{\label{fig2_iso} (Color online) Type-II Dirac points in PtSe$_2$. (a,b) Three-dimensional band structures in the (a) $k_x$-$k_z$ plane and (b) $k_z=k_z^c$ plane around the Dirac point. The bottom projection shows the iso-energy contour of electron and hole pocket with respect to $E_D$. (c,d,e) Three-dimensional iso-energy surface at (c) $E_F=E_D-30$ meV, (d) $E_D$, and (e) $E_D+30$ meV. Contact between electron and hole pockets occurs at the Dirac points (marked with blue arrows) when $E_F=E_D$.}
\end{figure}

From the above group theoretical analysis of the bands, we can clarify that type-II Dirac points exist in the PtSe$_2$ band structure. In order to have a direct visualization of the tilted Dirac cone, we plot the 3D band structure on the $k_y\!=\!0$ and $k_z\!=\!k_z^c$ planes around the Dirac point. As shown in Fig.~\ref{fig2_iso}(a,b), the two crossing bands exhibit linear dispersions in the vicinity of the Dirac point $D$ along both in-plane and out-of-plane directions. Moreover, the Dirac cone is strongly tilted along the $k_z$ direction, and consequently, a part of the upper cone becomes lower in energy than parts of the lower cone. This unique tilted Dirac cone, which is substantially different from the conventional ones in type-I Dirac semimetals, such as Na$_3$Bi and Cd$_3$As$_2$ \cite{Na3Bi,Cd3As2}, is similar to that of the recently discovered type-II Weyl fermion in WTe$_2$ and MoTe$_2$ \cite{type2Weyl,BinghaiMoTe2,*QPIMoTe2,*HsinLinMoWTe2,*ZJWangMoTe2,shuyunMoTe2,*DinghongMoTe2,*ZhouXJMoTe2,*YLChenMoTe2,*huang2016MoTe2,
wang2016spectroscopic,*wu2016observation,*bruno2016surface}. Therefore, the pair of Dirac points in PtSe$_2$ can be regarded as 3D type-II Dirac points. The existence of type-II Dirac points in PtSe$_2$ suggests that new kinds of quasi-particle, type-II 3D Dirac fermions which are different from the standard 3D Dirac fermions discovered before, can emerge in this material. Many physical properties of PtSe$_2$, such as magnetotransport anomalies, would be significantly different from those of type-I Dirac semimetals.

Next we study the Fermi surface topology and possible Lifshitz transition in PtSe$_2$ as the chemical potential varies.
Dirac semimetals were previously thought to have a point-like Fermi surface at the Dirac point and the Fermi surface expands to closed spheres or ellipsoids after $n$- or $p$-doping, which were observed in Na$_3$Bi and Cd$_3$As$_2$ by recent experiments \cite{Na3Bi_exp1,*Na3Bi_exp2,*Na3Bi_exp3,Cd3As2_exp1,*Cd3As2_exp2,*Cd3As2_exp3}. However, the Fermi surface of PtSe$_2$ shows quite different features when the chemical potential is fine tuned around the Dirac point.
The bottom projection of Fig.~\ref{fig2_iso}(a,b) shows the iso-energy contour of electron and hole pockets when the chemical potential is shifted to $E_D$. It is clear that the Dirac point appears as the contact point between electron and hole pockets in the $k_x$-$k_z$ plane (Fig.~\ref{fig2_iso}(a)). While the pockets shrink to the single Dirac point at the $k_z=k_z^c$ plane  (Fig.~\ref{fig2_iso}(b)). This can be also seen in the evolution of the 3D iso-energy surface as the chemical potential passes through $E_D$. 
As shown in Fig.~\ref{fig2_iso}(c), when $E_F<E_D$, there is a hole pocket with three-fold rotation symmetry at the center of the bulk BZ. The electron pockets are divided into two parts: The inner one consists of two half cones which disconnect with the hole pocket; whereas the outer one has the lantern shape surrounding other pockets. When the chemical potential increases to the energy of the Dirac point $E_F=E_D$, the hole pocket touches two inner electron pockets at the Dirac points as illustrated in Fig.~\ref{fig2_iso}(d). Further increasing the chemical potential disconnects the electron and hole pockets again, and the hole pocket vanishes gradually (see Fig.~\ref{fig2_iso}(e)). Evidently, the iso-energy surface evolves rapidly and undergoes a change in topology (i.e., Lifshitz transition) with tuning the chemical potential around the Dirac point. Therefore the Dirac point that emerges at the boundary between electron and hole pockets and possible Lifshitz transition on the Fermi surface can be readily observed when the chemical potential is brought down to the Dirac point by doping or other techniques \footnotemark[\value{footnote}].

To further reveal the nature of type-II Dirac points in PtSe$_2$, we derive the low-energy effective Hamiltonian using the theory of invariants\cite{bir1974symmetry} (see the Supplementary Materials \footnotemark[\value{footnote}] for more details). The first-principles calculations indicate that the wave functions of low-energy valence-band states at the $\Gamma$ point are mostly from the Se-$4p_{x,y,z}$ orbitals. Due to the $C_3$ rotation and inversion symmetries of the system, it is more convenient to make a linear combination of these orbitals as $|\eta, p_{x\pm iy},s\rangle$ and $|\eta,p_z,s\rangle$, where $\eta=\pm$ for bonding and antibonding states, $s=\uparrow\downarrow$ for spin.
By including spin-orbit coupling in the above atomic orbital picture, we need to consider the total angular momentum, and thus the new eigenstates at the $\Gamma$ point can be written as $|J=\frac{3}{2},J_z=\pm\frac{3}{2}\rangle$, $|J=\frac{3}{2},J_z=\pm\frac{1}{2}\rangle$ and $|J=\frac{1}{2},J_z=\pm\frac{1}{2}\rangle$.


At the $\Gamma$ point, the top two valence-band states belong to 2D IR $\Gamma_4^-$ and degenerate 1D IRs $\Gamma_5^-+\Gamma_6^-$, respectively (see Fig.~\ref{fig1_band}). The basis functions of these IRs are mainly composed of $|\frac{1}{2},\pm\frac{1}{2}\rangle$ and $|\frac{3}{2},\pm\frac{3}{2}\rangle$, according to their double group representations \cite{koster1963properties}.
In terms of the above four basis states (in the order of $|\frac{3}{2},\frac{3}{2}\rangle$, $|\frac{1}{2},-\frac{1}{2}\rangle$, $|\frac{3}{2},-\frac{3}{2}\rangle$ and $|\frac{1}{2},\frac{1}{2}\rangle$), we can construct an effective $4 \times 4$ Hamiltonian by considering the time-reversal, inversion and $D^3_{3d}$ symmetries\cite{Na3Bi}:
\begin{equation}
H_{\mathrm{eff}}(\mathbf{k})=\epsilon_0(\mathbf{k})+\left(
\begin{array}{cccc}
M(\mathbf{k}) & Ak_+ & 0 & B(\mathbf{k})\\
Ak_- & -M(\mathbf{k}) & -B(\mathbf{k}) & 0\\
0 & -B^*(\mathbf{k}) & M(\mathbf{k}) & Ak_-\\
B^*(\mathbf{k}) & 0 & Ak_+ & -M(\mathbf{k})\\
\end{array}
\right),\nonumber
\end{equation}
where $\epsilon_0(\mathbf{k}) = C_0+C_1k_z^2+C_2(k_x^2+k_y^2)$, $M(\mathbf{k})=M_0-M_1k_z^2-M_2(k_x^2+k_y^2)$, $B(\mathbf{k})=B_3k_zk^2_+$ and $k_{\pm}=k_x\pm ik_y$. The material-dependent parameters in the above Hamiltonian are determined by fitting the the energy spectrum of the effective Hamiltonian with that of DFT calculations. Our fitting yields $C_0\!=\!-0.2079$ eV, $C_1\!=\!-10.552$ eV\AA$^2$, $C_2\!=\!-12.134$ eV\AA$^2$, $M_0\!=\!-0.6362$ eV, $M_1\!=\!-6.8653$ eV\AA$^2$, $M_2\!=\!4.7005$ eV\AA$^2$ and $A\!=\!2.7602$ eV\AA. Evaluating the eigenvalues $E(\mathbf{k})=\epsilon_0(\mathbf{k})\pm\sqrt{M(\mathbf{k})^2+A^2k_+k_-+|B(\mathbf{k})|^2}$, we get a pair of gapless solutions at $\mathbf{k}^c=(0,0,k^c_z=\pm\sqrt{\frac{M_0}{M_1}})$, which are nothing but just the two Dirac points on the $k_z$ axis discussed above.

As we only concentrate on the vicinity of each crossing point $\mathbf{k}^c$, we choose $\mathbf{k}^c$ for reference and define $\mathbf{q}=\mathbf{k}-\mathbf{k}^c=(q_x,q_y,q_z)$. By expanding the Hamiltonian near the reference point and keeping only the linear order terms, we get
\begin{equation}
H^c(\mathbf{q})=\left(\begin{array}{cc}
h(\mathbf{q})&0\\
0&h^*(\mathbf{q})\\
\end{array}
\right)
\end{equation}
with
\begin{equation}
\begin{split}
h(\mathbf{q})&=q_z(2C_1k_z^c)\sigma_0\\
&+q_xA\sigma_x - q_yA\sigma_y-q_z(2M_1k_z^c)\sigma_z,
\end{split}
\end{equation}
where $\sigma_0$ is the unit matrix and $(\sigma_x, \sigma_y, \sigma_z)$ are Pauli matrices. The linearized Hamiltonian around the $\mathbf{k}^c$ point is nothing but the 3D anisotropic massless Dirac fermions, which can be decoupled into two Weyl fermions with degenerate energy. The energy spectrum is $\varepsilon_{\pm}(\mathbf{q})=T(\mathbf{q})\pm U(\mathbf{q})=2C_1k_z^c q_z\pm \sqrt{(Aq_x)^2+(Aq_y)^2+(2M_1k_z^cq_z)^2}$. Inserting the parameters obtained above into the formula, we can clearly see that $T(\hat{q}_z)>U(\hat{q}_z)$ along the $\hat{q}_z$ axis. Hence, the Dirac cone tilts strongly along the $\hat{k}_z$ direction, which causes the Dirac point to appear at the point where the open electron and hole pockets touch \cite{type2Weyl}. The Dirac points in PtSe$_2$ are therefore type-II Dirac points. As Weyl semimetals can, in principle, be realized by breaking either time-reversal or inversion symmetries of Dirac semimetals, type-II Weyl fermions are expected to be accessible in PtSe$_2$-class materials by splitting type-II Dirac points using magnetic doping or external strains.

\begin{figure}
\includegraphics[width =\columnwidth]{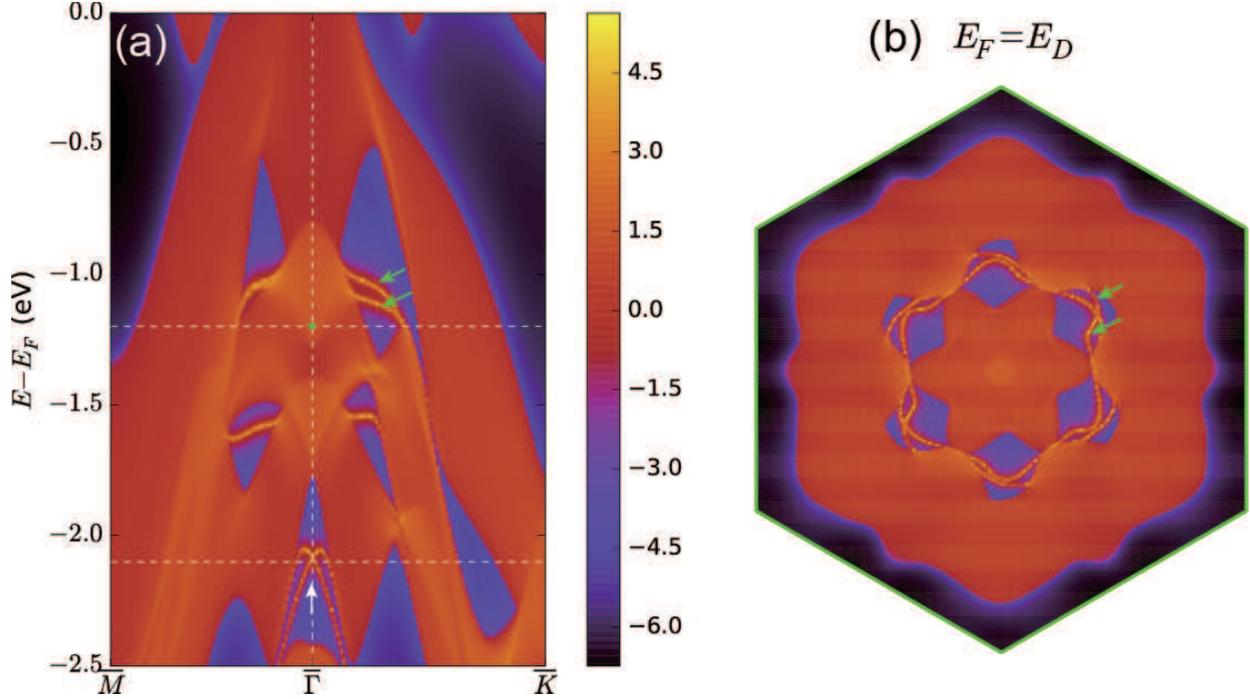}%
\caption{\label{fig4_surface} (Color online) The projected surface states and iso-energy surface for the (001) surface of PtSe$_2$.
(a) The projected surface density of states for the (001) surface. The projected Dirac points are denoted by green dots. The green arrows mark surface states around the projected Dirac point, whereas the white arrows represent the Dirac-cone surface state due to nontrivial $\mathbb{Z}_2$ topology.
(b) The iso-energy surface at $E_D$ of the (001) surface.}
\end{figure}

One of the most interesting phenomena of a Dirac semimetal is the presence of symmetry-protected surface states. We have computed the surface states for the (001) surface of PtSe$_2$, which are shown in Fig.~\ref{fig4_surface}. On the (001) surface, two Dirac points project to the $\bar{\Gamma}$ point of the surface BZ. According to previous analysis of conventional Dirac semimetals, such as Na$_3$Bi and Cd$_3$As$_2$ \cite{Na3Bi,Cd3As2}, there is just a single point on the Fermi surface of the (001) surface, and the signals of the surface bands depend sensitively on the surface potential
\cite{Na3Bi_exp1,Na3Bi_exp2,Na3Bi_exp3}. However, for the type-II Dirac points in PtSe$_2$, we do find some surface bands (marked with green arrows) between bulk bands at the (001) surface as shown in Fig.~\ref{fig4_surface}(a). By investigating the evolution of the surface states of PtSe$_2$ under uniaxial strain, we find that the two surface states start from continuous bulk states, disperse upward, and merge together at the projected Dirac point (for details of the calculation, see the Supplementary Materials \footnotemark[\value{footnote}]). The iso-energy surfaces at $E_F=E_D$ make it clear that the two
surface bands are located in the gapped circular region of the bulk-state continuum (Fig.~\ref{fig4_surface}(b)) These surface states, which are related to the symmetry-protected Dirac points, would be affected significantly when the trigonal symmetry is broken (see the Supplementary Materials \footnotemark[\value{footnote}]). In addition, due to the nontrivial $\mathbb{Z}_2$ topological gap between the $v_3$ and the $v_4$ valence bands, a Dirac-cone surface state lies deeply below the Fermi level around $-2.1$ eV and is well separated from bulk states in the (001) surface (see the white arrows in Fig.~\ref{fig4_surface}(a)). These surface states should be experimentally measurable by modern angle-resolved photoemission spectroscopy (ARPES) technique. In fact, similar novel surface states and type-II Dirac points of transition metal dichalcogendie PtTe$_2$ are observed in our recent ARPES measurements \cite{PtTe2}.

In conclusion, based on first-principles calculations and effective Hamiltonian analysis, we predict the existence of the type-II Dirac fermions that emerge at the boundary between electron and hole pockets in the transition metal dichalcogenide PtSe$_2$ class. A single pair of type-II Dirac points, which are protected by the trigonal rotation symmetry, exists in the $k_z$ axis of the bulk BZ. Moreover, novel surface states and doping-driven Lifshitz transition are observable in these materials. Our Rapid Communication provides useful guidance of experimentally detecting type-II Dirac fermions and related intriguing physical properties in a family of transition metal dichalcogenides.

\textit{Note}: In the final stages of preparing this Rapid Communication, we became aware of the independent work by Le \textit{et al}. \cite{le2016three} and Chang \textit{et al}. \cite{chang2016type} which reported similar type-II Dirac fermions in KMgBi and VAl$_3$ family materials, respectively.

\begin{acknowledgments}
We thank M. Yan for valuable discussions. This work was supported by the Ministry of Science and Technology of China (Grants No. 2016YFA0301001 and 2016YFA0301004) and the National Natural Science Foundation of China (Grant No. 11334006).
\end{acknowledgments}


\begin{thebibliography}{53}%
\makeatletter
\providecommand \@ifxundefined [1]{%
 \@ifx{#1\undefined}
}%
\providecommand \@ifnum [1]{%
 \ifnum #1\expandafter \@firstoftwo
 \else \expandafter \@secondoftwo
 \fi
}%
\providecommand \@ifx [1]{%
 \ifx #1\expandafter \@firstoftwo
 \else \expandafter \@secondoftwo
 \fi
}%
\providecommand \natexlab [1]{#1}%
\providecommand \enquote  [1]{``#1''}%
\providecommand \bibnamefont  [1]{#1}%
\providecommand \bibfnamefont [1]{#1}%
\providecommand \citenamefont [1]{#1}%
\providecommand \href@noop [0]{\@secondoftwo}%
\providecommand \href [0]{\begingroup \@sanitize@url \@href}%
\providecommand \@href[1]{\@@startlink{#1}\@@href}%
\providecommand \@@href[1]{\endgroup#1\@@endlink}%
\providecommand \@sanitize@url [0]{\catcode `\\12\catcode `\$12\catcode
  `\&12\catcode `\#12\catcode `\^12\catcode `\_12\catcode `\%12\relax}%
\providecommand \@@startlink[1]{}%
\providecommand \@@endlink[0]{}%
\providecommand \url  [0]{\begingroup\@sanitize@url \@url }%
\providecommand \@url [1]{\endgroup\@href {#1}{\urlprefix }}%
\providecommand \urlprefix  [0]{URL }%
\providecommand \Eprint [0]{\href }%
\providecommand \doibase [0]{http://dx.doi.org/}%
\providecommand \selectlanguage [0]{\@gobble}%
\providecommand \bibinfo  [0]{\@secondoftwo}%
\providecommand \bibfield  [0]{\@secondoftwo}%
\providecommand \translation [1]{[#1]}%
\providecommand \BibitemOpen [0]{}%
\providecommand \bibitemStop [0]{}%
\providecommand \bibitemNoStop [0]{.\EOS\space}%
\providecommand \EOS [0]{\spacefactor3000\relax}%
\providecommand \BibitemShut  [1]{\csname bibitem#1\endcsname}%
\let\auto@bib@innerbib\@empty
\bibitem [{\citenamefont {Wang}\ \emph {et~al.}(2012)\citenamefont {Wang},
  \citenamefont {Sun}, \citenamefont {Chen}, \citenamefont {Franchini},
  \citenamefont {Xu}, \citenamefont {Weng}, \citenamefont {Dai},\ and\
  \citenamefont {Fang}}]{Na3Bi}%
  \BibitemOpen
  \bibfield  {author} {\bibinfo {author} {\bibfnamefont {Z.}~\bibnamefont
  {Wang}}, \bibinfo {author} {\bibfnamefont {Y.}~\bibnamefont {Sun}}, \bibinfo
  {author} {\bibfnamefont {X.-Q.}\ \bibnamefont {Chen}}, \bibinfo {author}
  {\bibfnamefont {C.}~\bibnamefont {Franchini}}, \bibinfo {author}
  {\bibfnamefont {G.}~\bibnamefont {Xu}}, \bibinfo {author} {\bibfnamefont
  {H.}~\bibnamefont {Weng}}, \bibinfo {author} {\bibfnamefont {X.}~\bibnamefont
  {Dai}}, \ and\ \bibinfo {author} {\bibfnamefont {Z.}~\bibnamefont {Fang}},\
  }\href {\doibase 10.1103/PhysRevB.85.195320} {\bibfield  {journal} {\bibinfo
  {journal} {Phys. Rev. B}\ }\textbf {\bibinfo {volume} {85}},\ \bibinfo
  {pages} {195320} (\bibinfo {year} {2012})}\BibitemShut {NoStop}%
\bibitem [{\citenamefont {Wang}\ \emph {et~al.}(2013)\citenamefont {Wang},
  \citenamefont {Weng}, \citenamefont {Wu}, \citenamefont {Dai},\ and\
  \citenamefont {Fang}}]{Cd3As2}%
  \BibitemOpen
  \bibfield  {author} {\bibinfo {author} {\bibfnamefont {Z.}~\bibnamefont
  {Wang}}, \bibinfo {author} {\bibfnamefont {H.}~\bibnamefont {Weng}}, \bibinfo
  {author} {\bibfnamefont {Q.}~\bibnamefont {Wu}}, \bibinfo {author}
  {\bibfnamefont {X.}~\bibnamefont {Dai}}, \ and\ \bibinfo {author}
  {\bibfnamefont {Z.}~\bibnamefont {Fang}},\ }\href@noop {} {\bibfield
  {journal} {\bibinfo  {journal} {Phys. Rev. B}\ }\textbf {\bibinfo {volume}
  {88}},\ \bibinfo {pages} {125427} (\bibinfo {year} {2013})}\BibitemShut
  {NoStop}%
\bibitem [{\citenamefont {Wan}\ \emph {et~al.}(2011)\citenamefont {Wan},
  \citenamefont {Turner}, \citenamefont {Vishwanath},\ and\ \citenamefont
  {Savrasov}}]{WanXG}%
  \BibitemOpen
  \bibfield  {author} {\bibinfo {author} {\bibfnamefont {X.}~\bibnamefont
  {Wan}}, \bibinfo {author} {\bibfnamefont {A.~M.}\ \bibnamefont {Turner}},
  \bibinfo {author} {\bibfnamefont {A.}~\bibnamefont {Vishwanath}}, \ and\
  \bibinfo {author} {\bibfnamefont {S.~Y.}\ \bibnamefont {Savrasov}},\ }\href
  {\doibase 10.1103/PhysRevB.83.205101} {\bibfield  {journal} {\bibinfo
  {journal} {Phys. Rev. B}\ }\textbf {\bibinfo {volume} {83}},\ \bibinfo
  {pages} {205101} (\bibinfo {year} {2011})}\BibitemShut {NoStop}%
\bibitem [{\citenamefont {Burkov}\ and\ \citenamefont
  {Balents}(2011)}]{Balents}%
  \BibitemOpen
  \bibfield  {author} {\bibinfo {author} {\bibfnamefont {A.~A.}\ \bibnamefont
  {Burkov}}\ and\ \bibinfo {author} {\bibfnamefont {L.}~\bibnamefont
  {Balents}},\ }\href {\doibase 10.1103/PhysRevLett.107.127205} {\bibfield
  {journal} {\bibinfo  {journal} {Phys. Rev. Lett.}\ }\textbf {\bibinfo
  {volume} {107}},\ \bibinfo {pages} {127205} (\bibinfo {year}
  {2011})}\BibitemShut {NoStop}%
\bibitem [{\citenamefont {Weng}\ \emph {et~al.}(2015)\citenamefont {Weng},
  \citenamefont {Fang}, \citenamefont {Fang}, \citenamefont {Bernevig},\ and\
  \citenamefont {Dai}}]{TaAsPRX}%
  \BibitemOpen
  \bibfield  {author} {\bibinfo {author} {\bibfnamefont {H.}~\bibnamefont
  {Weng}}, \bibinfo {author} {\bibfnamefont {C.}~\bibnamefont {Fang}}, \bibinfo
  {author} {\bibfnamefont {Z.}~\bibnamefont {Fang}}, \bibinfo {author}
  {\bibfnamefont {B.~A.}\ \bibnamefont {Bernevig}}, \ and\ \bibinfo {author}
  {\bibfnamefont {X.}~\bibnamefont {Dai}},\ }\href {\doibase
  10.1103/PhysRevX.5.011029} {\bibfield  {journal} {\bibinfo  {journal} {Phys.
  Rev. X}\ }\textbf {\bibinfo {volume} {5}},\ \bibinfo {pages} {011029}
  (\bibinfo {year} {2015})}\BibitemShut {NoStop}%
\bibitem [{\citenamefont {Huang}\ \emph
  {et~al.}(2015{\natexlab{a}})\citenamefont {Huang}, \citenamefont {Xu},
  \citenamefont {Belopolski}, \citenamefont {Lee}, \citenamefont {Chang},
  \citenamefont {Wang}, \citenamefont {Alidoust}, \citenamefont {Bian},
  \citenamefont {Neupane}, \citenamefont {Zhang} \emph {et~al.}}]{TaAsHLin}%
  \BibitemOpen
  \bibfield  {author} {\bibinfo {author} {\bibfnamefont {S.-M.}\ \bibnamefont
  {Huang}}, \bibinfo {author} {\bibfnamefont {S.-Y.}\ \bibnamefont {Xu}},
  \bibinfo {author} {\bibfnamefont {I.}~\bibnamefont {Belopolski}}, \bibinfo
  {author} {\bibfnamefont {C.-C.}\ \bibnamefont {Lee}}, \bibinfo {author}
  {\bibfnamefont {G.}~\bibnamefont {Chang}}, \bibinfo {author} {\bibfnamefont
  {B.}~\bibnamefont {Wang}}, \bibinfo {author} {\bibfnamefont {N.}~\bibnamefont
  {Alidoust}}, \bibinfo {author} {\bibfnamefont {G.}~\bibnamefont {Bian}},
  \bibinfo {author} {\bibfnamefont {M.}~\bibnamefont {Neupane}}, \bibinfo
  {author} {\bibfnamefont {C.}~\bibnamefont {Zhang}},  \emph {et~al.},\
  }\href@noop {} {\bibfield  {journal} {\bibinfo  {journal} {Nat. Commun.}\
  }\textbf {\bibinfo {volume} {6}} (\bibinfo {year}
  {2015}{\natexlab{a}})}\BibitemShut {NoStop}%
\bibitem [{\citenamefont {Liu}\ and\ \citenamefont
  {Vanderbilt}(2014)}]{Jianpeng}%
  \BibitemOpen
  \bibfield  {author} {\bibinfo {author} {\bibfnamefont {J.}~\bibnamefont
  {Liu}}\ and\ \bibinfo {author} {\bibfnamefont {D.}~\bibnamefont
  {Vanderbilt}},\ }\href {\doibase 10.1103/PhysRevB.90.155316} {\bibfield
  {journal} {\bibinfo  {journal} {Phys. Rev. B}\ }\textbf {\bibinfo {volume}
  {90}},\ \bibinfo {pages} {155316} (\bibinfo {year} {2014})}\BibitemShut
  {NoStop}%
\bibitem [{\citenamefont {Cao}\ \emph {et~al.}(2016)\citenamefont {Cao},
  \citenamefont {Tang}, \citenamefont {Zhang}, \citenamefont {Duan},\ and\
  \citenamefont {Rubio}}]{CaoWendong}%
  \BibitemOpen
  \bibfield  {author} {\bibinfo {author} {\bibfnamefont {W.}~\bibnamefont
  {Cao}}, \bibinfo {author} {\bibfnamefont {P.}~\bibnamefont {Tang}}, \bibinfo
  {author} {\bibfnamefont {S.-C.}\ \bibnamefont {Zhang}}, \bibinfo {author}
  {\bibfnamefont {W.}~\bibnamefont {Duan}}, \ and\ \bibinfo {author}
  {\bibfnamefont {A.}~\bibnamefont {Rubio}},\ }\href@noop {} {\bibfield
  {journal} {\bibinfo  {journal} {Phys. Rev. B}\ }\textbf {\bibinfo {volume}
  {93}},\ \bibinfo {pages} {241117} (\bibinfo {year} {2016})}\BibitemShut
  {NoStop}%
\bibitem [{\citenamefont {Nielsen}\ and\ \citenamefont
  {Ninomiya}(1983)}]{ChiralAnomaly}%
  \BibitemOpen
  \bibfield  {author} {\bibinfo {author} {\bibfnamefont {H.~B.}\ \bibnamefont
  {Nielsen}}\ and\ \bibinfo {author} {\bibfnamefont {M.}~\bibnamefont
  {Ninomiya}},\ }\href@noop {} {\bibfield  {journal} {\bibinfo  {journal}
  {Phys. Lett. B}\ }\textbf {\bibinfo {volume} {130}},\ \bibinfo {pages} {389}
  (\bibinfo {year} {1983})}\BibitemShut {NoStop}%
\bibitem [{\citenamefont {Abrikosov}(1998)}]{QMR}%
  \BibitemOpen
  \bibfield  {author} {\bibinfo {author} {\bibfnamefont {A.~A.}\ \bibnamefont
  {Abrikosov}},\ }\href {\doibase 10.1103/PhysRevB.58.2788} {\bibfield
  {journal} {\bibinfo  {journal} {Phys. Rev. B}\ }\textbf {\bibinfo {volume}
  {58}},\ \bibinfo {pages} {2788} (\bibinfo {year} {1998})}\BibitemShut
  {NoStop}%
\bibitem [{\citenamefont {Soluyanov}\ \emph {et~al.}(2015)\citenamefont
  {Soluyanov}, \citenamefont {Gresch}, \citenamefont {Wang}, \citenamefont
  {Wu}, \citenamefont {Troyer}, \citenamefont {Dai},\ and\ \citenamefont
  {Bernevig}}]{type2Weyl}%
  \BibitemOpen
  \bibfield  {author} {\bibinfo {author} {\bibfnamefont {A.~A.}\ \bibnamefont
  {Soluyanov}}, \bibinfo {author} {\bibfnamefont {D.}~\bibnamefont {Gresch}},
  \bibinfo {author} {\bibfnamefont {Z.}~\bibnamefont {Wang}}, \bibinfo {author}
  {\bibfnamefont {Q.}~\bibnamefont {Wu}}, \bibinfo {author} {\bibfnamefont
  {M.}~\bibnamefont {Troyer}}, \bibinfo {author} {\bibfnamefont
  {X.}~\bibnamefont {Dai}}, \ and\ \bibinfo {author} {\bibfnamefont {B.~A.}\
  \bibnamefont {Bernevig}},\ }\href@noop {} {\bibfield  {journal} {\bibinfo
  {journal} {Nature}\ }\textbf {\bibinfo {volume} {527}},\ \bibinfo {pages}
  {495} (\bibinfo {year} {2015})}\BibitemShut {NoStop}%
\bibitem [{\citenamefont {Sun}\ \emph {et~al.}(2015)\citenamefont {Sun},
  \citenamefont {Wu}, \citenamefont {Ali}, \citenamefont {Felser},\ and\
  \citenamefont {Yan}}]{BinghaiMoTe2}%
  \BibitemOpen
  \bibfield  {author} {\bibinfo {author} {\bibfnamefont {Y.}~\bibnamefont
  {Sun}}, \bibinfo {author} {\bibfnamefont {S.-C.}\ \bibnamefont {Wu}},
  \bibinfo {author} {\bibfnamefont {M.~N.}\ \bibnamefont {Ali}}, \bibinfo
  {author} {\bibfnamefont {C.}~\bibnamefont {Felser}}, \ and\ \bibinfo {author}
  {\bibfnamefont {B.}~\bibnamefont {Yan}},\ }\href {\doibase
  10.1103/PhysRevB.92.161107} {\bibfield  {journal} {\bibinfo  {journal} {Phys.
  Rev. B}\ }\textbf {\bibinfo {volume} {92}},\ \bibinfo {pages} {161107}
  (\bibinfo {year} {2015})}\BibitemShut {NoStop}%
\bibitem [{\citenamefont {Kourtis}\ \emph {et~al.}(2016)\citenamefont
  {Kourtis}, \citenamefont {Li}, \citenamefont {Wang}, \citenamefont
  {Yazdani},\ and\ \citenamefont {Bernevig}}]{QPIMoTe2}%
  \BibitemOpen
  \bibfield  {author} {\bibinfo {author} {\bibfnamefont {S.}~\bibnamefont
  {Kourtis}}, \bibinfo {author} {\bibfnamefont {J.}~\bibnamefont {Li}},
  \bibinfo {author} {\bibfnamefont {Z.}~\bibnamefont {Wang}}, \bibinfo {author}
  {\bibfnamefont {A.}~\bibnamefont {Yazdani}}, \ and\ \bibinfo {author}
  {\bibfnamefont {B.~A.}\ \bibnamefont {Bernevig}},\ }\href {\doibase
  10.1103/PhysRevB.93.041109} {\bibfield  {journal} {\bibinfo  {journal} {Phys.
  Rev. B}\ }\textbf {\bibinfo {volume} {93}},\ \bibinfo {pages} {041109}
  (\bibinfo {year} {2016})}\BibitemShut {NoStop}%
\bibitem [{\citenamefont {Chang}\ \emph
  {et~al.}(2016{\natexlab{a}})\citenamefont {Chang}, \citenamefont {Xu},
  \citenamefont {Chang} \emph {et~al.}}]{HsinLinMoWTe2}%
  \BibitemOpen
  \bibfield  {author} {\bibinfo {author} {\bibfnamefont {T.-R.}\ \bibnamefont
  {Chang}}, \bibinfo {author} {\bibfnamefont {S.-Y.}\ \bibnamefont {Xu}},
  \bibinfo {author} {\bibfnamefont {G.}~\bibnamefont {Chang}},  \emph
  {et~al.},\ }\href@noop {} {\bibfield  {journal} {\bibinfo  {journal} {Nat.
  Commun.}\ }\textbf {\bibinfo {volume} {7}} (\bibinfo {year}
  {2016}{\natexlab{a}})}\BibitemShut {NoStop}%
\bibitem [{\citenamefont {Wang}\ \emph
  {et~al.}(2016{\natexlab{a}})\citenamefont {Wang}, \citenamefont {Gresch},
  \citenamefont {Soluyanov} \emph {et~al.}}]{ZJWangMoTe2}%
  \BibitemOpen
  \bibfield  {author} {\bibinfo {author} {\bibfnamefont {Z.}~\bibnamefont
  {Wang}}, \bibinfo {author} {\bibfnamefont {D.}~\bibnamefont {Gresch}},
  \bibinfo {author} {\bibfnamefont {A.~A.}\ \bibnamefont {Soluyanov}},  \emph
  {et~al.},\ }\href {\doibase 10.1103/PhysRevLett.117.056805} {\bibfield
  {journal} {\bibinfo  {journal} {Phys. Rev. Lett.}\ }\textbf {\bibinfo
  {volume} {117}},\ \bibinfo {pages} {056805} (\bibinfo {year}
  {2016}{\natexlab{a}})}\BibitemShut {NoStop}%
\bibitem [{\citenamefont {Deng}\ \emph {et~al.}(2016)\citenamefont {Deng},
  \citenamefont {Wan}, \citenamefont {Deng} \emph {et~al.}}]{shuyunMoTe2}%
  \BibitemOpen
  \bibfield  {author} {\bibinfo {author} {\bibfnamefont {K.}~\bibnamefont
  {Deng}}, \bibinfo {author} {\bibfnamefont {G.}~\bibnamefont {Wan}}, \bibinfo
  {author} {\bibfnamefont {P.}~\bibnamefont {Deng}},  \emph {et~al.},\ }\href
  {\doibase 10.1038/nphys3871} {\bibfield  {journal} {\bibinfo  {journal} {Nat.
  Phys.}\ } (\bibinfo {year} {2016}),\ 10.1038/nphys3871}\BibitemShut {NoStop}%
\bibitem [{\citenamefont {Xu}\ \emph {et~al.}(2016)\citenamefont {Xu},
  \citenamefont {Wang}, \citenamefont {Weber} \emph {et~al.}}]{DinghongMoTe2}%
  \BibitemOpen
  \bibfield  {author} {\bibinfo {author} {\bibfnamefont {N.}~\bibnamefont
  {Xu}}, \bibinfo {author} {\bibfnamefont {Z.}~\bibnamefont {Wang}}, \bibinfo
  {author} {\bibfnamefont {A.}~\bibnamefont {Weber}},  \emph {et~al.},\
  }\href@noop {} {\bibfield  {journal} {\bibinfo  {journal} {arXiv preprint
  arXiv:1604.02116}\ } (\bibinfo {year} {2016})}\BibitemShut {NoStop}%
\bibitem [{\citenamefont {Liang}\ \emph {et~al.}(2016)\citenamefont {Liang},
  \citenamefont {Huang}, \citenamefont {Nie} \emph {et~al.}}]{ZhouXJMoTe2}%
  \BibitemOpen
  \bibfield  {author} {\bibinfo {author} {\bibfnamefont {A.}~\bibnamefont
  {Liang}}, \bibinfo {author} {\bibfnamefont {J.}~\bibnamefont {Huang}},
  \bibinfo {author} {\bibfnamefont {S.}~\bibnamefont {Nie}},  \emph {et~al.},\
  }\href@noop {} {\bibfield  {journal} {\bibinfo  {journal} {arXiv preprint
  arXiv:1604.01706}\ } (\bibinfo {year} {2016})}\BibitemShut {NoStop}%
\bibitem [{\citenamefont {Jiang}\ \emph {et~al.}(2016)\citenamefont {Jiang},
  \citenamefont {Liu}, \citenamefont {Sun} \emph {et~al.}}]{YLChenMoTe2}%
  \BibitemOpen
  \bibfield  {author} {\bibinfo {author} {\bibfnamefont {J.}~\bibnamefont
  {Jiang}}, \bibinfo {author} {\bibfnamefont {Z.}~\bibnamefont {Liu}}, \bibinfo
  {author} {\bibfnamefont {Y.}~\bibnamefont {Sun}},  \emph {et~al.},\
  }\href@noop {} {\bibfield  {journal} {\bibinfo  {journal} {arXiv preprint
  arXiv:1604.00139}\ } (\bibinfo {year} {2016})}\BibitemShut {NoStop}%
\bibitem [{\citenamefont {Huang}\ \emph
  {et~al.}(2016{\natexlab{a}})\citenamefont {Huang}, \citenamefont {McCormick},
  \citenamefont {Ochi} \emph {et~al.}}]{huang2016MoTe2}%
  \BibitemOpen
  \bibfield  {author} {\bibinfo {author} {\bibfnamefont {L.}~\bibnamefont
  {Huang}}, \bibinfo {author} {\bibfnamefont {T.~M.}\ \bibnamefont
  {McCormick}}, \bibinfo {author} {\bibfnamefont {M.}~\bibnamefont {Ochi}},
  \emph {et~al.},\ }\href@noop {} {\bibfield  {journal} {\bibinfo  {journal}
  {arXiv preprint arXiv:1603.06482}\ } (\bibinfo {year}
  {2016}{\natexlab{a}})}\BibitemShut {NoStop}%
\bibitem [{\citenamefont {Wang}\ \emph
  {et~al.}(2016{\natexlab{b}})\citenamefont {Wang}, \citenamefont {Zhang},
  \citenamefont {Huang} \emph {et~al.}}]{wang2016spectroscopic}%
  \BibitemOpen
  \bibfield  {author} {\bibinfo {author} {\bibfnamefont {C.}~\bibnamefont
  {Wang}}, \bibinfo {author} {\bibfnamefont {Y.}~\bibnamefont {Zhang}},
  \bibinfo {author} {\bibfnamefont {J.}~\bibnamefont {Huang}},  \emph
  {et~al.},\ }\href@noop {} {\bibfield  {journal} {\bibinfo  {journal} {arXiv
  preprint arXiv:1604.04218}\ } (\bibinfo {year}
  {2016}{\natexlab{b}})}\BibitemShut {NoStop}%
\bibitem [{\citenamefont {Wu}\ \emph {et~al.}(2016)\citenamefont {Wu},
  \citenamefont {Jo}, \citenamefont {Mou} \emph {et~al.}}]{wu2016observation}%
  \BibitemOpen
  \bibfield  {author} {\bibinfo {author} {\bibfnamefont {Y.}~\bibnamefont
  {Wu}}, \bibinfo {author} {\bibfnamefont {N.~H.}\ \bibnamefont {Jo}}, \bibinfo
  {author} {\bibfnamefont {D.}~\bibnamefont {Mou}},  \emph {et~al.},\
  }\href@noop {} {\bibfield  {journal} {\bibinfo  {journal} {arXiv preprint
  arXiv:1604.05176}\ } (\bibinfo {year} {2016})}\BibitemShut {NoStop}%
\bibitem [{\citenamefont {Bruno}\ \emph {et~al.}(2016)\citenamefont {Bruno},
  \citenamefont {Tamai}, \citenamefont {Wu} \emph {et~al.}}]{bruno2016surface}%
  \BibitemOpen
  \bibfield  {author} {\bibinfo {author} {\bibfnamefont {F.}~\bibnamefont
  {Bruno}}, \bibinfo {author} {\bibfnamefont {A.}~\bibnamefont {Tamai}},
  \bibinfo {author} {\bibfnamefont {Q.}~\bibnamefont {Wu}},  \emph {et~al.},\
  }\href@noop {} {\bibfield  {journal} {\bibinfo  {journal} {arXiv preprint
  arXiv:1604.02411}\ } (\bibinfo {year} {2016})}\BibitemShut {NoStop}%
\bibitem [{\citenamefont {Volovik}(2016)}]{volovik2016lifshitz}%
  \BibitemOpen
  \bibfield  {author} {\bibinfo {author} {\bibfnamefont {G.}~\bibnamefont
  {Volovik}},\ }\href@noop {} {\bibfield  {journal} {\bibinfo  {journal} {arXiv
  preprint arXiv:1604.00849}\ } (\bibinfo {year} {2016})}\BibitemShut {NoStop}%
\bibitem [{\citenamefont {Xu}\ \emph {et~al.}(2015{\natexlab{a}})\citenamefont
  {Xu}, \citenamefont {Zhang},\ and\ \citenamefont {Zhang}}]{YongXu}%
  \BibitemOpen
  \bibfield  {author} {\bibinfo {author} {\bibfnamefont {Y.}~\bibnamefont
  {Xu}}, \bibinfo {author} {\bibfnamefont {F.}~\bibnamefont {Zhang}}, \ and\
  \bibinfo {author} {\bibfnamefont {C.}~\bibnamefont {Zhang}},\ }\href@noop {}
  {\bibfield  {journal} {\bibinfo  {journal} {Phys. Rev. Lett.}\ }\textbf
  {\bibinfo {volume} {115}},\ \bibinfo {pages} {265304} (\bibinfo {year}
  {2015}{\natexlab{a}})}\BibitemShut {NoStop}%
\bibitem [{\citenamefont {Liu}\ \emph {et~al.}(2014{\natexlab{a}})\citenamefont
  {Liu}, \citenamefont {Zhou}, \citenamefont {Zhang} \emph
  {et~al.}}]{Na3Bi_exp1}%
  \BibitemOpen
  \bibfield  {author} {\bibinfo {author} {\bibfnamefont {Z.~K.}\ \bibnamefont
  {Liu}}, \bibinfo {author} {\bibfnamefont {B.}~\bibnamefont {Zhou}}, \bibinfo
  {author} {\bibfnamefont {Y.}~\bibnamefont {Zhang}},  \emph {et~al.},\ }\href
  {\doibase 10.1126/science.1245085} {\bibfield  {journal} {\bibinfo  {journal}
  {Science}\ }\textbf {\bibinfo {volume} {343}},\ \bibinfo {pages} {864}
  (\bibinfo {year} {2014}{\natexlab{a}})}\BibitemShut {NoStop}%
\bibitem [{\citenamefont {Xu}\ \emph {et~al.}(2015{\natexlab{b}})\citenamefont
  {Xu}, \citenamefont {Liu}, \citenamefont {Kushwaha} \emph
  {et~al.}}]{Na3Bi_exp2}%
  \BibitemOpen
  \bibfield  {author} {\bibinfo {author} {\bibfnamefont {S.-Y.}\ \bibnamefont
  {Xu}}, \bibinfo {author} {\bibfnamefont {C.}~\bibnamefont {Liu}}, \bibinfo
  {author} {\bibfnamefont {S.~K.}\ \bibnamefont {Kushwaha}},  \emph {et~al.},\
  }\href {\doibase 10.1126/science.1256742} {\bibfield  {journal} {\bibinfo
  {journal} {Science}\ }\textbf {\bibinfo {volume} {347}},\ \bibinfo {pages}
  {294} (\bibinfo {year} {2015}{\natexlab{b}})}\BibitemShut {NoStop}%
\bibitem [{\citenamefont {Xiong}\ \emph {et~al.}(2015)\citenamefont {Xiong},
  \citenamefont {Kushwaha}, \citenamefont {Liang} \emph {et~al.}}]{Na3Bi_exp3}%
  \BibitemOpen
  \bibfield  {author} {\bibinfo {author} {\bibfnamefont {J.}~\bibnamefont
  {Xiong}}, \bibinfo {author} {\bibfnamefont {S.~K.}\ \bibnamefont {Kushwaha}},
  \bibinfo {author} {\bibfnamefont {T.}~\bibnamefont {Liang}},  \emph
  {et~al.},\ }\href {\doibase 10.1126/science.aac6089} {\bibfield  {journal}
  {\bibinfo  {journal} {Science}\ }\textbf {\bibinfo {volume} {350}},\ \bibinfo
  {pages} {413} (\bibinfo {year} {2015})}\BibitemShut {NoStop}%
\bibitem [{\citenamefont {Neupane}\ \emph {et~al.}(2014)\citenamefont
  {Neupane}, \citenamefont {Xu}, \citenamefont {Sankar} \emph
  {et~al.}}]{Cd3As2_exp1}%
  \BibitemOpen
  \bibfield  {author} {\bibinfo {author} {\bibfnamefont {M.}~\bibnamefont
  {Neupane}}, \bibinfo {author} {\bibfnamefont {S.-Y.}\ \bibnamefont {Xu}},
  \bibinfo {author} {\bibfnamefont {R.}~\bibnamefont {Sankar}},  \emph
  {et~al.},\ }\href {http://dx.doi.org/10.1038/ncomms4786} {\bibfield
  {journal} {\bibinfo  {journal} {Nat. Commun.}\ }\textbf {\bibinfo {volume}
  {5}} (\bibinfo {year} {2014})}\BibitemShut {NoStop}%
\bibitem [{\citenamefont {Liu}\ \emph {et~al.}(2014{\natexlab{b}})\citenamefont
  {Liu}, \citenamefont {Jiang}, \citenamefont {Zhou} \emph
  {et~al.}}]{Cd3As2_exp2}%
  \BibitemOpen
  \bibfield  {author} {\bibinfo {author} {\bibfnamefont {Z.~K.}\ \bibnamefont
  {Liu}}, \bibinfo {author} {\bibfnamefont {J.}~\bibnamefont {Jiang}}, \bibinfo
  {author} {\bibfnamefont {B.}~\bibnamefont {Zhou}},  \emph {et~al.},\ }\href
  {http://dx.doi.org/10.1038/nmat3990} {\bibfield  {journal} {\bibinfo
  {journal} {Nat. Mater.}\ }\textbf {\bibinfo {volume} {13}},\ \bibinfo {pages}
  {677} (\bibinfo {year} {2014}{\natexlab{b}})}\BibitemShut {NoStop}%
\bibitem [{\citenamefont {Borisenko}\ \emph {et~al.}(2014)\citenamefont
  {Borisenko}, \citenamefont {Gibson}, \citenamefont {Evtushinsky} \emph
  {et~al.}}]{Cd3As2_exp3}%
  \BibitemOpen
  \bibfield  {author} {\bibinfo {author} {\bibfnamefont {S.}~\bibnamefont
  {Borisenko}}, \bibinfo {author} {\bibfnamefont {Q.}~\bibnamefont {Gibson}},
  \bibinfo {author} {\bibfnamefont {D.}~\bibnamefont {Evtushinsky}},  \emph
  {et~al.},\ }\href {\doibase 10.1103/PhysRevLett.113.027603} {\bibfield
  {journal} {\bibinfo  {journal} {Phys. Rev. Lett.}\ }\textbf {\bibinfo
  {volume} {113}},\ \bibinfo {pages} {027603} (\bibinfo {year}
  {2014})}\BibitemShut {NoStop}%
\bibitem [{\citenamefont {Perdew}\ \emph {et~al.}(1996)\citenamefont {Perdew},
  \citenamefont {Burke},\ and\ \citenamefont {Ernzerhof}}]{PBE}%
  \BibitemOpen
  \bibfield  {author} {\bibinfo {author} {\bibfnamefont {J.~P.}\ \bibnamefont
  {Perdew}}, \bibinfo {author} {\bibfnamefont {K.}~\bibnamefont {Burke}}, \
  and\ \bibinfo {author} {\bibfnamefont {M.}~\bibnamefont {Ernzerhof}},\
  }\href@noop {} {\bibfield  {journal} {\bibinfo  {journal} {Phys.\ Rev.
  Lett.}\ }\textbf {\bibinfo {volume} {77}},\ \bibinfo {pages} {3865} (\bibinfo
  {year} {1996})}\BibitemShut {NoStop}%
\bibitem [{\citenamefont {Kresse}\ and\ \citenamefont
  {Furthm\"{u}ller}(1996)}]{VASP}%
  \BibitemOpen
  \bibfield  {author} {\bibinfo {author} {\bibfnamefont {G.}~\bibnamefont
  {Kresse}}\ and\ \bibinfo {author} {\bibfnamefont {J.}~\bibnamefont
  {Furthm\"{u}ller}},\ }\href@noop {} {\bibfield  {journal} {\bibinfo
  {journal} {Comput. Mater. Sci.}\ }\textbf {\bibinfo {volume} {6}},\ \bibinfo
  {pages} {15} (\bibinfo {year} {1996})}\BibitemShut {NoStop}%
\bibitem [{\citenamefont {Marzari}\ and\ \citenamefont
  {Vanderbilt}(1997)}]{wannier1}%
  \BibitemOpen
  \bibfield  {author} {\bibinfo {author} {\bibfnamefont {N.}~\bibnamefont
  {Marzari}}\ and\ \bibinfo {author} {\bibfnamefont {D.}~\bibnamefont
  {Vanderbilt}},\ }\href {\doibase 10.1103/PhysRevB.56.12847} {\bibfield
  {journal} {\bibinfo  {journal} {Phys. Rev. B}\ }\textbf {\bibinfo {volume}
  {56}},\ \bibinfo {pages} {12847} (\bibinfo {year} {1997})}\BibitemShut
  {NoStop}%
\bibitem [{\citenamefont {Souza}\ \emph {et~al.}(2001)\citenamefont {Souza},
  \citenamefont {Marzari},\ and\ \citenamefont {Vanderbilt}}]{wannier2}%
  \BibitemOpen
  \bibfield  {author} {\bibinfo {author} {\bibfnamefont {I.}~\bibnamefont
  {Souza}}, \bibinfo {author} {\bibfnamefont {N.}~\bibnamefont {Marzari}}, \
  and\ \bibinfo {author} {\bibfnamefont {D.}~\bibnamefont {Vanderbilt}},\
  }\href {\doibase 10.1103/PhysRevB.65.035109} {\bibfield  {journal} {\bibinfo
  {journal} {Phys. Rev. B}\ }\textbf {\bibinfo {volume} {65}},\ \bibinfo
  {pages} {035109} (\bibinfo {year} {2001})}\BibitemShut {NoStop}%
\bibitem [{\citenamefont {Mostofi}\ \emph {et~al.}(2008)\citenamefont
  {Mostofi}, \citenamefont {Yates}, \citenamefont {Lee}, \citenamefont {Souza},
  \citenamefont {Vanderbilt},\ and\ \citenamefont {Marzari}}]{wannier90}%
  \BibitemOpen
  \bibfield  {author} {\bibinfo {author} {\bibfnamefont {A.~A.}\ \bibnamefont
  {Mostofi}}, \bibinfo {author} {\bibfnamefont {J.~R.}\ \bibnamefont {Yates}},
  \bibinfo {author} {\bibfnamefont {Y.-S.}\ \bibnamefont {Lee}}, \bibinfo
  {author} {\bibfnamefont {I.}~\bibnamefont {Souza}}, \bibinfo {author}
  {\bibfnamefont {D.}~\bibnamefont {Vanderbilt}}, \ and\ \bibinfo {author}
  {\bibfnamefont {N.}~\bibnamefont {Marzari}},\ }\href@noop {} {\bibfield
  {journal} {\bibinfo  {journal} {Comput. Phys. Commun.}\ }\textbf {\bibinfo
  {volume} {178}},\ \bibinfo {pages} {685} (\bibinfo {year}
  {2008})}\BibitemShut {NoStop}%
\bibitem [{\citenamefont {Huang}\ \emph
  {et~al.}(2015{\natexlab{b}})\citenamefont {Huang}, \citenamefont {Wang},
  \citenamefont {Luo}, \citenamefont {Liu}, \citenamefont {L\"u}, \citenamefont
  {Wu},\ and\ \citenamefont {Duan}}]{InterfaceTICI}%
  \BibitemOpen
  \bibfield  {author} {\bibinfo {author} {\bibfnamefont {H.}~\bibnamefont
  {Huang}}, \bibinfo {author} {\bibfnamefont {Z.}~\bibnamefont {Wang}},
  \bibinfo {author} {\bibfnamefont {N.}~\bibnamefont {Luo}}, \bibinfo {author}
  {\bibfnamefont {Z.}~\bibnamefont {Liu}}, \bibinfo {author} {\bibfnamefont
  {R.}~\bibnamefont {L\"u}}, \bibinfo {author} {\bibfnamefont {J.}~\bibnamefont
  {Wu}}, \ and\ \bibinfo {author} {\bibfnamefont {W.}~\bibnamefont {Duan}},\
  }\href {\doibase 10.1103/PhysRevB.92.075138} {\bibfield  {journal} {\bibinfo
  {journal} {Phys. Rev. B}\ }\textbf {\bibinfo {volume} {92}},\ \bibinfo
  {pages} {075138} (\bibinfo {year} {2015}{\natexlab{b}})}\BibitemShut
  {NoStop}%
\bibitem [{\citenamefont {Huang}\ \emph
  {et~al.}(2016{\natexlab{b}})\citenamefont {Huang}, \citenamefont {Liu},
  \citenamefont {Vanderbilt},\ and\ \citenamefont {Duan}}]{huanghq_NodalLine}%
  \BibitemOpen
  \bibfield  {author} {\bibinfo {author} {\bibfnamefont {H.}~\bibnamefont
  {Huang}}, \bibinfo {author} {\bibfnamefont {J.}~\bibnamefont {Liu}}, \bibinfo
  {author} {\bibfnamefont {D.}~\bibnamefont {Vanderbilt}}, \ and\ \bibinfo
  {author} {\bibfnamefont {W.}~\bibnamefont {Duan}},\ }\href {\doibase
  10.1103/PhysRevB.93.201114} {\bibfield  {journal} {\bibinfo  {journal} {Phys.
  Rev. B}\ }\textbf {\bibinfo {volume} {93}},\ \bibinfo {pages} {201114}
  (\bibinfo {year} {2016}{\natexlab{b}})}\BibitemShut {NoStop}%
\bibitem [{\citenamefont {L{\'o}pez~Sancho}\ \emph {et~al.}(1984)\citenamefont
  {L{\'o}pez~Sancho}, \citenamefont {L{\'o}pez~Sancho},\ and\ \citenamefont
  {Rubio}}]{lopez}%
  \BibitemOpen
  \bibfield  {author} {\bibinfo {author} {\bibfnamefont {M.~P.}\ \bibnamefont
  {L{\'o}pez~Sancho}}, \bibinfo {author} {\bibfnamefont {J.~M.}\ \bibnamefont
  {L{\'o}pez~Sancho}}, \ and\ \bibinfo {author} {\bibfnamefont
  {J.}~\bibnamefont {Rubio}},\ }\href@noop {} {\bibfield  {journal} {\bibinfo
  {journal} {J. Phys. F}\ }\textbf {\bibinfo {volume} {14}},\ \bibinfo {pages}
  {1205} (\bibinfo {year} {1984})}\BibitemShut {NoStop}%
\bibitem [{\citenamefont {L{\'o}pez~Sancho}\ \emph {et~al.}(1985)\citenamefont
  {L{\'o}pez~Sancho}, \citenamefont {L{\'o}pez~Sancho},\ and\ \citenamefont
  {Rubio}}]{lopez2}%
  \BibitemOpen
  \bibfield  {author} {\bibinfo {author} {\bibfnamefont {M.~P.}\ \bibnamefont
  {L{\'o}pez~Sancho}}, \bibinfo {author} {\bibfnamefont {J.~M.}\ \bibnamefont
  {L{\'o}pez~Sancho}}, \ and\ \bibinfo {author} {\bibfnamefont
  {J.}~\bibnamefont {Rubio}},\ }\href
  {http://stacks.iop.org/0305-4608/15/i=4/a=009} {\bibfield  {journal}
  {\bibinfo  {journal} {J. Phys. F}\ }\textbf {\bibinfo {volume} {15}},\
  \bibinfo {pages} {851} (\bibinfo {year} {1985})}\BibitemShut {NoStop}%
\bibitem [{\citenamefont {Huang}\ \emph {et~al.}(2014)\citenamefont {Huang},
  \citenamefont {Liu},\ and\ \citenamefont {Duan}}]{PhysRevB.90.195105}%
  \BibitemOpen
  \bibfield  {author} {\bibinfo {author} {\bibfnamefont {H.}~\bibnamefont
  {Huang}}, \bibinfo {author} {\bibfnamefont {J.}~\bibnamefont {Liu}}, \ and\
  \bibinfo {author} {\bibfnamefont {W.}~\bibnamefont {Duan}},\ }\href {\doibase
  10.1103/PhysRevB.90.195105} {\bibfield  {journal} {\bibinfo  {journal} {Phys.
  Rev. B}\ }\textbf {\bibinfo {volume} {90}},\ \bibinfo {pages} {195105}
  (\bibinfo {year} {2014})}\BibitemShut {NoStop}%
\bibitem [{\citenamefont {Huang}\ \emph
  {et~al.}(2015{\natexlab{c}})\citenamefont {Huang}, \citenamefont {Liu},
  \citenamefont {Zhang}, \citenamefont {Duan},\ and\ \citenamefont
  {Vanderbilt}}]{semiDirac}%
  \BibitemOpen
  \bibfield  {author} {\bibinfo {author} {\bibfnamefont {H.}~\bibnamefont
  {Huang}}, \bibinfo {author} {\bibfnamefont {Z.}~\bibnamefont {Liu}}, \bibinfo
  {author} {\bibfnamefont {H.}~\bibnamefont {Zhang}}, \bibinfo {author}
  {\bibfnamefont {W.}~\bibnamefont {Duan}}, \ and\ \bibinfo {author}
  {\bibfnamefont {D.}~\bibnamefont {Vanderbilt}},\ }\href {\doibase
  10.1103/PhysRevB.92.161115} {\bibfield  {journal} {\bibinfo  {journal} {Phys.
  Rev. B}\ }\textbf {\bibinfo {volume} {92}},\ \bibinfo {pages} {161115}
  (\bibinfo {year} {2015}{\natexlab{c}})}\BibitemShut {NoStop}%
\bibitem [{\citenamefont {Bradlyn}\ \emph {et~al.}(2016)\citenamefont
  {Bradlyn}, \citenamefont {Cano}, \citenamefont {Wang}, \citenamefont {Cava},\
  and\ \citenamefont {Bernevig}}]{newfermion}%
  \BibitemOpen
  \bibfield  {author} {\bibinfo {author} {\bibfnamefont {B.}~\bibnamefont
  {Bradlyn}}, \bibinfo {author} {\bibfnamefont {J.}~\bibnamefont {Cano}},
  \bibinfo {author} {\bibfnamefont {Z.}~\bibnamefont {Wang}}, \bibinfo {author}
  {\bibfnamefont {R.}~\bibnamefont {Cava}}, \ and\ \bibinfo {author}
  {\bibfnamefont {B.~A.}\ \bibnamefont {Bernevig}},\ }\href {\doibase
  10.1126/science.aaf5037} {\bibfield  {journal} {\bibinfo  {journal}
  {Science}\ }\textbf {\bibinfo {volume} {353}},\ \bibinfo {pages} {6299}
  (\bibinfo {year} {2016})}\BibitemShut {NoStop}%
\bibitem [{\citenamefont {Weng}\ \emph {et~al.}(2016)\citenamefont {Weng},
  \citenamefont {Fang}, \citenamefont {Fang},\ and\ \citenamefont
  {Dai}}]{triple1}%
  \BibitemOpen
  \bibfield  {author} {\bibinfo {author} {\bibfnamefont {H.}~\bibnamefont
  {Weng}}, \bibinfo {author} {\bibfnamefont {C.}~\bibnamefont {Fang}}, \bibinfo
  {author} {\bibfnamefont {Z.}~\bibnamefont {Fang}}, \ and\ \bibinfo {author}
  {\bibfnamefont {X.}~\bibnamefont {Dai}},\ }\href {\doibase
  10.1103/PhysRevB.93.241202} {\bibfield  {journal} {\bibinfo  {journal} {Phys.
  Rev. B}\ }\textbf {\bibinfo {volume} {93}},\ \bibinfo {pages} {241202}
  (\bibinfo {year} {2016})}\BibitemShut {NoStop}%
\bibitem [{\citenamefont {Zhu}\ \emph {et~al.}(2016)\citenamefont {Zhu},
  \citenamefont {Winkler}, \citenamefont {Wu}, \citenamefont {Li},\ and\
  \citenamefont {Soluyanov}}]{triple2}%
  \BibitemOpen
  \bibfield  {author} {\bibinfo {author} {\bibfnamefont {Z.}~\bibnamefont
  {Zhu}}, \bibinfo {author} {\bibfnamefont {G.~W.}\ \bibnamefont {Winkler}},
  \bibinfo {author} {\bibfnamefont {Q.}~\bibnamefont {Wu}}, \bibinfo {author}
  {\bibfnamefont {J.}~\bibnamefont {Li}}, \ and\ \bibinfo {author}
  {\bibfnamefont {A.~A.}\ \bibnamefont {Soluyanov}},\ }\href {\doibase
  10.1103/PhysRevX.6.031003} {\bibfield  {journal} {\bibinfo  {journal} {Phys.
  Rev. X}\ }\textbf {\bibinfo {volume} {6}},\ \bibinfo {pages} {031003}
  (\bibinfo {year} {2016})}\BibitemShut {NoStop}%
\bibitem [{\citenamefont {Koster}(1963)}]{koster1963properties}%
  \BibitemOpen
  \bibfield  {author} {\bibinfo {author} {\bibfnamefont {G.~F.}\ \bibnamefont
  {Koster}},\ }\href@noop {} {\emph {\bibinfo {title} {Properties of the
  thirty-two point groups}}},\ Vol.~\bibinfo {volume} {24}\ (\bibinfo
  {publisher} {The MIT Press},\ \bibinfo {address} {Cambridge, Massachusetts},\
  \bibinfo {year} {1963})\BibitemShut {NoStop}%
\bibitem [{Note1()}]{Note1}%
  \BibitemOpen
  \bibinfo {note} {\label {fn}See Supplemental Material at http://link.aps.org/supplemental/
10.1103/PhysRevB.94.121117 for Type-II Dirac fermions in the
PtSe2 class of transition metal dichalcogenides.}\BibitemShut
  {Stop}%
\bibitem [{\citenamefont {Soluyanov}\ and\ \citenamefont
  {Vanderbilt}(2011)}]{alexey2}%
  \BibitemOpen
  \bibfield  {author} {\bibinfo {author} {\bibfnamefont {A.~A.}\ \bibnamefont
  {Soluyanov}}\ and\ \bibinfo {author} {\bibfnamefont {D.}~\bibnamefont
  {Vanderbilt}},\ }\href {\doibase 10.1103/PhysRevB.83.235401} {\bibfield
  {journal} {\bibinfo  {journal} {Phys. Rev. B}\ }\textbf {\bibinfo {volume}
  {83}},\ \bibinfo {pages} {235401} (\bibinfo {year} {2011})}\BibitemShut
  {NoStop}%
\bibitem [{\citenamefont {Liu}\ \emph {et~al.}(2015)\citenamefont {Liu},
  \citenamefont {Zhao}, \citenamefont {Li}, \citenamefont {Lin}, \citenamefont
  {Liang}, \citenamefont {Hu}, \citenamefont {Ding}, \citenamefont {Xu},
  \citenamefont {He}, \citenamefont {Zhao} \emph {et~al.}}]{pdte2}%
  \BibitemOpen
  \bibfield  {author} {\bibinfo {author} {\bibfnamefont {Y.}~\bibnamefont
  {Liu}}, \bibinfo {author} {\bibfnamefont {J.-Z.}\ \bibnamefont {Zhao}},
  \bibinfo {author} {\bibfnamefont {Y.}~\bibnamefont {Li}}, \bibinfo {author}
  {\bibfnamefont {C.-T.}\ \bibnamefont {Lin}}, \bibinfo {author} {\bibfnamefont
  {A.-J.}\ \bibnamefont {Liang}}, \bibinfo {author} {\bibfnamefont
  {C.}~\bibnamefont {Hu}}, \bibinfo {author} {\bibfnamefont {Y.}~\bibnamefont
  {Ding}}, \bibinfo {author} {\bibfnamefont {Y.}~\bibnamefont {Xu}}, \bibinfo
  {author} {\bibfnamefont {S.-L.}\ \bibnamefont {He}}, \bibinfo {author}
  {\bibfnamefont {L.}~\bibnamefont {Zhao}},  \emph {et~al.},\ }\href@noop {}
  {\bibfield  {journal} {\bibinfo  {journal} {Chin. Phys. Lett.}\ }\textbf
  {\bibinfo {volume} {32}},\ \bibinfo {pages} {067303} (\bibinfo {year}
  {2015})}\BibitemShut {NoStop}%
\bibitem [{\citenamefont {Bir}\ and\ \citenamefont
  {Pikus}(1974)}]{bir1974symmetry}%
  \BibitemOpen
  \bibfield  {author} {\bibinfo {author} {\bibfnamefont {G.~L.}\ \bibnamefont
  {Bir}}\ and\ \bibinfo {author} {\bibfnamefont {G.~E.}\ \bibnamefont
  {Pikus}},\ }\href@noop {} {\emph {\bibinfo {title} {Symmetry and
  strain-induced effects in semiconductors}}},\ Vol.\ \bibinfo {volume} {624}\
  (\bibinfo  {publisher} {Wiley, New York},\ \bibinfo {year}
  {1974})\BibitemShut {NoStop}%
\bibitem [{\citenamefont {Yan}\ \emph {et~al.}(2016)\citenamefont {Yan},
  \citenamefont {Huang}, \citenamefont {Zhang}, \citenamefont {Wang},
  \citenamefont {Yao}, \citenamefont {Deng}, \citenamefont {Wan}, \citenamefont
  {Zhang}, \citenamefont {Arita}, \citenamefont {Yang}, \citenamefont {Sun},
  \citenamefont {Yao}, \citenamefont {Wu}, \citenamefont {Fan}, \citenamefont
  {Duan},\ and\ \citenamefont {Zhou}}]{PtTe2}%
  \BibitemOpen
  \bibfield  {author} {\bibinfo {author} {\bibfnamefont {M.}~\bibnamefont
  {Yan}}, \bibinfo {author} {\bibfnamefont {H.}~\bibnamefont {Huang}}, \bibinfo
  {author} {\bibfnamefont {K.}~\bibnamefont {Zhang}}, \bibinfo {author}
  {\bibfnamefont {E.}~\bibnamefont {Wang}}, \bibinfo {author} {\bibfnamefont
  {W.}~\bibnamefont {Yao}}, \bibinfo {author} {\bibfnamefont {K.}~\bibnamefont
  {Deng}}, \bibinfo {author} {\bibfnamefont {G.}~\bibnamefont {Wan}}, \bibinfo
  {author} {\bibfnamefont {H.}~\bibnamefont {Zhang}}, \bibinfo {author}
  {\bibfnamefont {M.}~\bibnamefont {Arita}}, \bibinfo {author} {\bibfnamefont
  {H.}~\bibnamefont {Yang}}, \bibinfo {author} {\bibfnamefont {Z.}~\bibnamefont
  {Sun}}, \bibinfo {author} {\bibfnamefont {H.}~\bibnamefont {Yao}}, \bibinfo
  {author} {\bibfnamefont {Y.}~\bibnamefont {Wu}}, \bibinfo {author}
  {\bibfnamefont {S.}~\bibnamefont {Fan}}, \bibinfo {author} {\bibfnamefont
  {W.}~\bibnamefont {Duan}}, \ and\ \bibinfo {author} {\bibfnamefont
  {S.}~\bibnamefont {Zhou}},\ }\href@noop {} {\bibfield  {journal} {\bibinfo
  {journal} {arXiv preprint arXiv:1607.03643}\ } (\bibinfo {year}
  {2016})}\BibitemShut {NoStop}%
\bibitem [{\citenamefont {Le}\ \emph {et~al.}(2016)\citenamefont {Le},
  \citenamefont {Qin}, \citenamefont {Wu}, \citenamefont {Dai}, \citenamefont
  {Fu},\ and\ \citenamefont {Hu}}]{le2016three}%
  \BibitemOpen
  \bibfield  {author} {\bibinfo {author} {\bibfnamefont {C.}~\bibnamefont
  {Le}}, \bibinfo {author} {\bibfnamefont {S.}~\bibnamefont {Qin}}, \bibinfo
  {author} {\bibfnamefont {X.}~\bibnamefont {Wu}}, \bibinfo {author}
  {\bibfnamefont {X.}~\bibnamefont {Dai}}, \bibinfo {author} {\bibfnamefont
  {P.}~\bibnamefont {Fu}}, \ and\ \bibinfo {author} {\bibfnamefont
  {J.}~\bibnamefont {Hu}},\ }\href@noop {} {\bibfield  {journal} {\bibinfo
  {journal} {arXiv preprint arXiv:1606.05042}\ } (\bibinfo {year}
  {2016})}\BibitemShut {NoStop}%
\bibitem [{\citenamefont {Chang}\ \emph
  {et~al.}(2016{\natexlab{b}})\citenamefont {Chang}, \citenamefont {Xu},
  \citenamefont {Sanchez}, \citenamefont {Huang}, \citenamefont {Chang},
  \citenamefont {Hsu}, \citenamefont {Bian}, \citenamefont {Belopolski},
  \citenamefont {Yu}, \citenamefont {Xu} \emph {et~al.}}]{chang2016type}%
  \BibitemOpen
  \bibfield  {author} {\bibinfo {author} {\bibfnamefont {T.-R.}\ \bibnamefont
  {Chang}}, \bibinfo {author} {\bibfnamefont {S.-Y.}\ \bibnamefont {Xu}},
  \bibinfo {author} {\bibfnamefont {D.~S.}\ \bibnamefont {Sanchez}}, \bibinfo
  {author} {\bibfnamefont {S.-M.}\ \bibnamefont {Huang}}, \bibinfo {author}
  {\bibfnamefont {G.}~\bibnamefont {Chang}}, \bibinfo {author} {\bibfnamefont
  {C.-H.}\ \bibnamefont {Hsu}}, \bibinfo {author} {\bibfnamefont
  {G.}~\bibnamefont {Bian}}, \bibinfo {author} {\bibfnamefont {I.}~\bibnamefont
  {Belopolski}}, \bibinfo {author} {\bibfnamefont {Z.-M.}\ \bibnamefont {Yu}},
  \bibinfo {author} {\bibfnamefont {X.}~\bibnamefont {Xu}},  \emph {et~al.},\
  }\href@noop {} {\bibfield  {journal} {\bibinfo  {journal} {arXiv preprint
  arXiv:1606.07555}\ } (\bibinfo {year} {2016}{\natexlab{b}})}\BibitemShut
  {NoStop}%
\end{thebibliography}
\providecommand{\noopsort}[1]{}\providecommand{\singleletter}[1]{#1}%

\end{document}